# Social Networks as a Tool for a Higher Education Institution Image Creation

Olha Anisimova[1] [0000-0002-8016-9361], Valeriia Vasylenko[2][0000-0002-2370-5615], Solomia Fedushko[3] [0000-0001-7548-5856]

[1]Vasyl' Stus Donetsk National University, Vinnytsia, Ukraine
[3]Lviv Polytechnic National University, Ukraine

o.anisimova@donnu.edu.ua[1], v.vasilenko@donnu.edu.ua[2],
solomiia.s.fedushko@lpnu.ua[3]

**Abstract.** The article presents the dynamics of social networks users increase, depending on the total world population from 2010 to 2018. It also identifies the most popular social networks in Ukraine. The systematic risk indicator of using social networks relative to the total number of Internet resources users is determined. Types of social intercourse in the process of the higher education institution image creation are presented. The peculiarities of using social networks in the formation of a positive image of an educational institution are highlighted. The statistical indicators of user actions in the official group of the Faculty of Mathematics and Information Technologies of Vasyl' Stus Donetsk National University in January, February and March 2019 are presented, as well as the average attraction coefficient of users depending on the subject of publications. The main technologies of astroturfing in the creation process of the higher education institution negative image are considered.

## 1      Introduction and Motivation

There is a rapid increase of the global Internet user numbers in the XXI century. First of all, this is due to the convenience and practicality of using a variety of tools for searching pertinent information, storage of data in cloud technologies, creation and support of communications, operative news reporting, etc. Social network is one of such tool that provides the ability to create private virtual space.
The analytical agency "Statista" represents the data demonstrating that number of social networks users has reached 3,196 billion people (in 2019), with the age group of the majority falling from 16 to 24 years old [22]. In other words, it can be assumed that the vast majority of users are schoolchildren and graduates of educational institutions that are enrollees and constitute a significant part of the target audience of the institution of higher education (IHE).

It leads to the formation and maintenance of a quality information policy using information resources of modern social networks in order to create a positive image for university enrollees, establishing and maintaining partnerships with similar educational and scientific institutions comes to the fore.

## 2    Related Research

A large number of scholars engaged in the formation of the image of a higher education institution. N. V. Horbenko [8] considers the need to introduce image management in the marketing system of the modern university; the problem of a IHE positive image creation by conducting a quality information policy is considered by S. M. Pavlov [15, 16], R. Korzh, A. Peleshchyshyn, S. Fedushko [10], Y. Syerov [12]; the problem of universities image creation through the prism of the image components itself and the signs of student satisfaction are engaged by N. Azoury, L. Daou, C. E. Khoury [1].

The use of social networks as an influence tool on behavior, the activity of the IHE target audience are considered by Ukrainian and foreign researches. G. I Batychko, O. R. Veliyeva [2] consider the using impact of the IHE Internet representation in social networks. A. Toda, R. Carmo, A. Silva, I. Bittencourt, S. Isotani [23] explores the use of social networks in educational contexts in order to increase of consumers' productivity, engage and motivation. M. Vitoropoulou and V. Karyotis [25], on the one hand, define the possibilities of tracking the distribution of various kinds of information in social sources; on the other hand, examine the characteristics of using the metrics of social networks analysis.

O. S. Petrenko [18], D. Holland, A. Krause, J. Provencher, T. Seltzer [7] are worth noticing. Authors consider technologies specifics of positive and negative influence of social networks on public opinion.

The peculiarities of the astroturfing technologies as the creation of artificial public opinion are considered by J. Peng, S. Detchon, K. R. Choo, H. Ashman [17]. R. Korzh, A. Peleshchyshyn, S. Fedushko, Y. Syerov [11] investigate possible methods for protecting the university's information image from short-term and ongoing aggressive actions in social networks.

However, it should be noted that there are no studies that fully disclose and comprehensively characterize the specifics of the use of social networks in the IHE image creation process. It determines the purpose of this study.

The purpose of the present research is to determine the peculiarities of the use of social networks in the process of higher education institution image creation.

## 3    Methodology of research

The methodology involves the number of general scientific and special methods of cognition. In particular, there are the following: the method of analysis and systematization of scientific literature, logical method, monitoring, observation and method of research results visualization.

## 4  Basic Points Statement

There is no single interpretation of the term «image» in modern science. Let's single out one of the most accurate definitions of the image in the era of modern information days. It defines an image as a communicative unit of influence on mass consciousness [9].

Conducting an analogy with the process of forming the image of a higher education institution, the consciousness control of a particular target audience, which can include enrollees and their parents, partner institutions, competitors, employers, and state could take place.

Two main components of the image have to be used in order to create a holistic image of the IHE. The information component includes a set of all representations (knowledge) about the institution itself: history, values, traditions of the IHE, peculiarities of its functioning. The estimated component includes a general idea of the educational institution characteristics and determines the relationship of the audience to the institution itself. The second one involves the features of its emotional perception, the formation of the readiness of the audience to act in different ways, consciously relying on those rules and regulations that operate in an educational institution [5, 15].

Social networks are one of the modern tool for providing the ability to establish communication links and work with the audience consciousness from all social and communication technologies.

Social networks are defined as a platform, an online service or a website created for social relationship formation, reflection and organization. [6].

The popularity of social networks is beyond doubt. Thus, according to the analytical agency "Statista", the number of Internet users in 2019 reached 4,021 billion people, almost 66 percent of which are social networks users.

The dynamics of the social networks users increase, depending on the total population in the world from 2010 to 2018, is presented in Fig. 1. [13, 14, 19].

To establish the necessity of using the information resources of social networks in the process of forming the image of the institution of higher education, we will determine the dependence, the coefficient of sensitivity (reliability) of the number of users of social networks.

First of all, use the formula for determining the systematic risk:

$$\beta = \frac{V_{R_i R}}{\sigma_R^2} \qquad (1)$$

where $R$ – a random variable characterizing the entire economy; $R_i$ – a random variable characterizing a particular industry.

It should be noted that the industry with the indicator:

- $\beta = 1$ has a fluctuation of results equaled to the market;
- $\beta < 1$ – less marketable;
- $\beta > 1$ – more marketable.

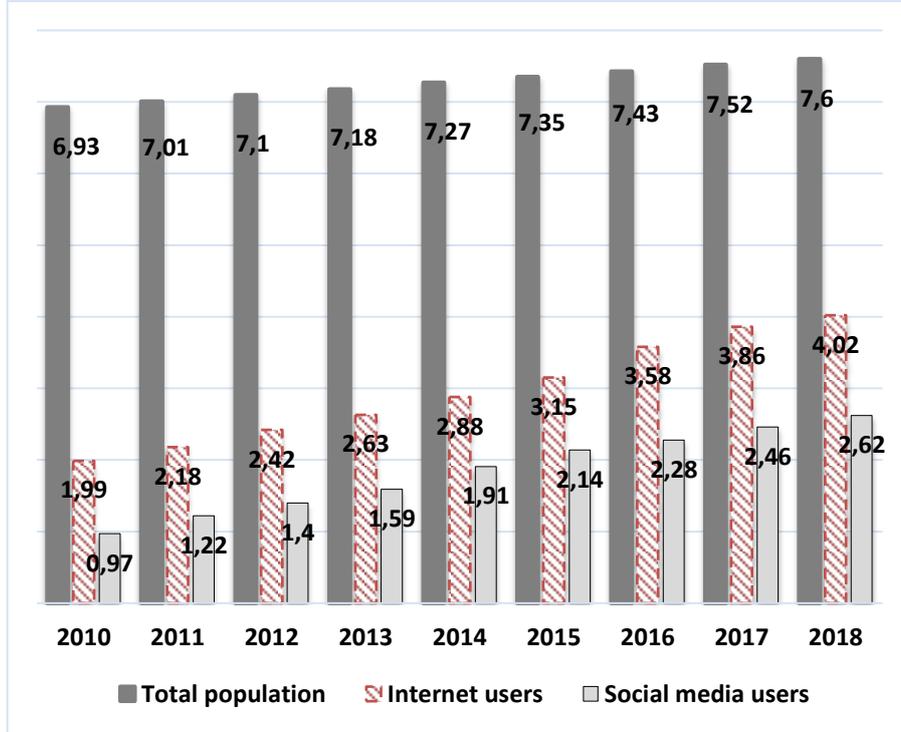

**Fig. 1.** Statistics on the use of social networks in the world from 2010 to 2018 (in billion).

The higher $\beta$ means the higher industry risk [20].

We will carry out the calculations on three stages:

I. Determination of the sensitivity coefficient of the number of Internet resources users (Ri = y) relative to the total population in the world (R = z).

II. Determination of the sensitivity coefficient of the number of social networks users (Ri = x) relative to the total population in the world (R = z).

III. Determination of the coefficient of sensitivity of the number of social networks users (Ri = x) relative to users of Internet resources in general (R = y).

The coefficient of sensitivity (reliability) of the number of social networks users is determined by the formula:

$$\beta = \frac{V_{xy}}{\sigma_y^2}, \qquad (2)$$

where

$$V_{xy} = \overline{xy} - \bar{x} \cdot \bar{y} \qquad (3)$$

and

$$\sigma_y^2 = \overline{y^2} - (\bar{y})^2. \qquad (4)$$

The sensitivity ratio of the number of Internet resources users relative to the total population in the world ($\beta_1$) is 3,11. The sensitivity ratio of the number of social networks users relative to the total population in the world ($\beta_2$) is 2,51. $\beta_3$ is the coefficient of sensitivity of the number of social networks users in relation to users of Internet resources in general is 0,77. It means the systematic risk index $\beta_3$ less than 1. In this case it is possible to conclude the following: the use of information resources of social networks is a prerequisite for the formation of the image of the IHE in the modern information and communication space.

If we consider the Ukrainian audience, then more than 11 million people are active users of Facebook. Instagram and Youtube take the second and the third place respectively (Fig. 2) [27].

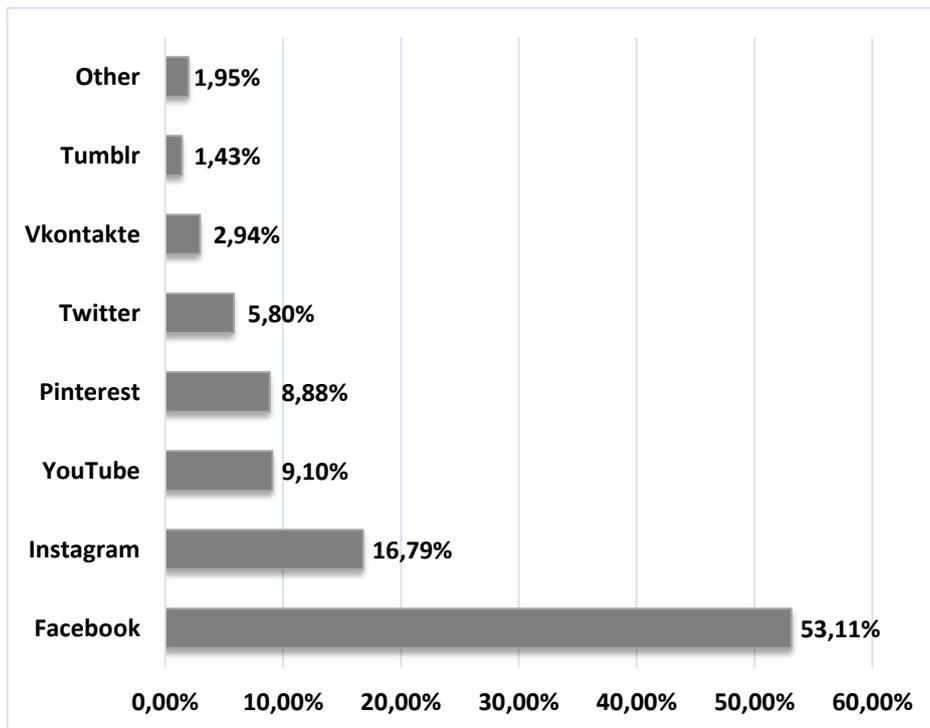

**Fig. 2.** The most popular Ukrainian social networks in 2018.

Consequently, it concludes that social networks are the tool affected the consciousness of population because of virtue of its prevalence and popularity.

One of the peculiarities of using social networks is the possibility of forming communications, creating certain social connections.

It is possible to highlight such social connections in shaping the image of a higher education institution (Fig. 3).

Fig. 2 shows strong social connections formed between the participants in the communication process permanent connected between themselves and the institution of higher education, using, besides social networks, other communication technologies (e-mail, mobile communication).

Latent social ties are formed between those target audience members who have potentially possible links, but have not yet established them.

Social networks allow ones to activate latent connections and transfer the members of the target audience to the role of «acquaintances». In other words, there is a transition to weak social ties [24].

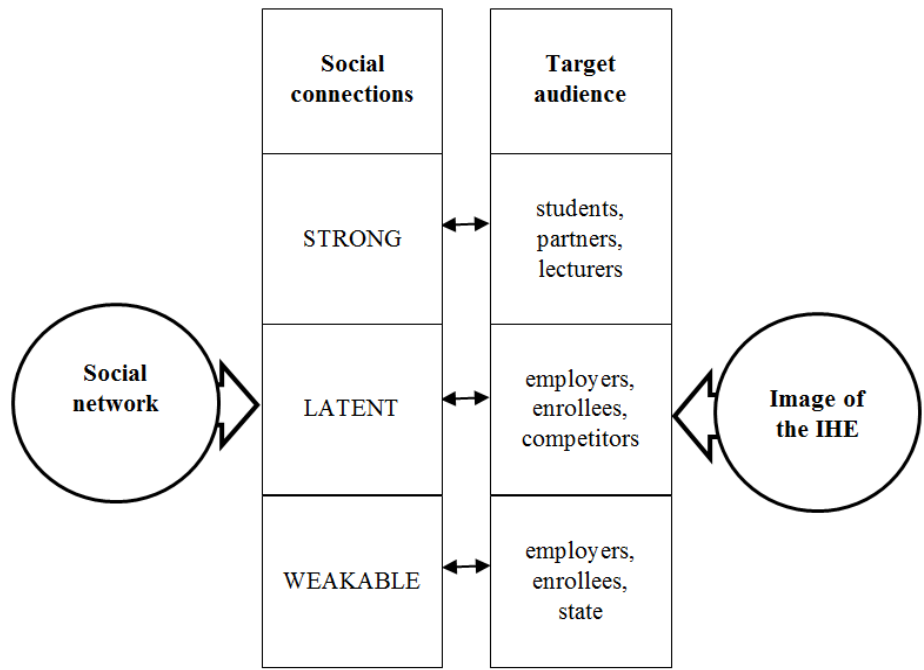

**Fig. 3.** Types of social connections during the image of the IHE creation.

Determine the features of using social networks during the image of IHE creation on their functional features (see Tab. 1) [21]. In such a manner, information resources of social networks provide IHE the following ability:

— to target users allocating the target audience by tracking the actions of the group members: subscribing to the group, "posting" the publications, placing the likes / dislikes;
— determination of information requests of potential enrollees and implementation of operational information services for users: holding consultations in personal correspondence, providing public answers to questions in the comments;
— create and support business relations with the target audience.

To determine the most popular content among Facebook social network users, it is possible to calculate the average engagement rate.

**Table 1.** Features of using social networks during the image of IHE creation.

| Social network function | Aspect of the IHE image creation | Information resources of social networks |
|---|---|---|
| Communication | Establishing business contacts with the target audience. Possibility of operative information exchange between the participants of the educational process (students, teachers) | Official groups of educational institutions, faculty, faculty in Facebook, Telegram, acounts in Instagram. Social Networks for Scholars: Google Scholar, Academia.edu, ResearchGate. Messengers for answering requests, «posting» publications, adding comments to them |
| Informational | Determination of information requests of enrollees. Creating information services for users. Publication of information on actual problems of the educational process | Subscriptions and group calls, individual publications. Consultations of enrollees, students, other participants of the educational process in public and / or personal correspondence. Comments of publications by group members |
| Socializing | Overcoming the communication barrier, obstacles in communicating with classmates, and teachers. Formation of a wide range of participants for communication | Virtual communication (text, audio, video messages) in personal correspondence. Adding friends and teachers to friends |
| Identification | Comparison of different IHE, students who study in them. Selection of the target audience: search for potential enrollees, partner institutions | Searching IHE, users with predefined parameters: age, gender, country; search for belonging to a particular group |
| Entertaining | Attracting the target audience. Ability to realize creative abilities | Publication of photo-, audio-, video-records of cultural events held in the IHE in official accounts, the accounts of the IHE themselves, «transferring» enrollees by students, teachers to their personal pages |

For this purpose, let's analyze all publications of the Faculty of Mathematics and Information Technologies (FMIT) of Vasyl' Stus Donetsk National University official group in the period from January to March in 2019 [4]. Three main topics of published records are defined:

1. «Festive» publications: birthday greetings of employees and students of the faculty; conducting, celebrating of festive events.
2. Thematic publications about employees and students: participation in educational events (olympiads, conferences), vocational guidance sessions, schedule of training sessions, examination sessions, participation in professionals' workshops.
3. University publications related to the organization of the educational process: the selection of students' free-choice courses; calls for study abroad; internships opportunities; job vacancies.

The table presents the main statistical performance of users, depending on the topics of publications in the official group of the Faculty of Mathematics and Information Technologies of Donetsk National University named after Vasyl' Stus during the defined period.

Table 2. Statistic representation of users in the official FMIT group on Facebook.

| The type of publications on Facebook | Number of users in the group | Number of publications | Involvement | |
|---|---|---|---|---|
| | | | Clicks for publication | Reactions-likes, comments, reposts |
| «Festive» publications | 610 | 9 | 546 | 315 |
| Thematic publications | | 7 | 371 | 151 |
| University publications | | 11 | 193 | 66 |

The formula for determining the average engagement rate for users is the follow [26]:

$$\text{Avrage Engagement Rate (AER)} == \frac{\text{Engagement Volume (Likes + Clicks + Comments + Reposts)} \div \textit{Number of posts}}{\text{Number of users}} \times 100\% \quad (5)$$

After conducting the calculations, it can be determined that the most popular among the group's subscribers are the publications on celebratory themes, AER which made up 15,7%. The second place takes thematic publications, the average rate of attraction of which was 12,2%. The smallest popularity in general university records, where the AER was 3,9%, although by statistic their number is the largest.

In the such way, the main features of using social networks in the creation of a positive image of higher education institution is identified. However, it should be noted that social networks are only at first glance a transparent instrument, objective and free of ideologies. In fact, social networks can be used to generate artificial public opinion that could mean the creation of an astroturfing process.

Under astroturfing we will mean «using modern software, or specially hired paid users for artificial management of public opinion». The main goal of astroturfing is to crowd out the thoughts of real people on web forums, to organize counterfeit online campaigns that create the impression that a large number of people require something specific or oppose something [3].

Consider the basic technologies of astroturfing implemented in social networks for the formation of a negative image of the institution of higher education [18]:

1. Invited comments and «posts» on Internet resources.
2. For example, competitor institutions may leave comments under publications in the official IHE group, indicating false information that the institution closes or the institution has not get a license for the activity.
3. Prepaid trolling including posting messages of provocative, humiliating content in order to provoke a controversy, to indulge the indignation, the rage of opponents.
4. For example, posting a post that professor of a IHE faculty, require bribes, from students for successful assignments, examinations or dissemination of information about an «intimate» scandal with the participation of an adult teacher and minor student.
5. Creating pages of «clones» or «fake» pages.
6. Social networks allow the free creation of pages of «artificial» personalities which false information disseminated through personal posts, commentary on publications; distribution of publications with a mark on other social network users.
7. The placement of «viral» photographic, audio and video materials in public access and distribution among users of the social network.

More effective is the placement of such materials in the «top», the most popular groups, or «promotion» (artificial increase) of likes, comments on such publications.

Consequently, the use of the above mentioned technologies of social networks can provide an opportunity to form an «artificial» negative public opinion about a particular institution of higher education, thereby reducing its rating and spoiling the overall impression. The change in the behavior of the target audience will be observed as follows: entrants and their parents will not have the will and in future to join the submitted IHE; potential employers, other educational and scientific institutions do not want to establish business relations, and students will be deducted or transferred to other departments / other institutions of higher education.

As we see, the use of social networks in the process of forming the image of a higher education institution is a very interesting and innovative direction that needs further research.

## 5      Conclusions

The dynamics of increase of social networks users on the total population and Internet users in the world from 2010 till 2018 is presented. The necessity of using information resources of social networks in the creation of the image of the higher education institution in the modern information and communication space due to the determination of the indicator of the systematic risk of using social networks in relation to the total number of users of Internet resources has been proved. Three main types of social connections are considered in shaping the image of a higher education institution depending on its target audience. The aspects of the higher education institution positive image in the functional use of information resources of social networks are described. An analysis of all publications of the Faculty of Mathematics and Information Technologies of the Vasyl' Stus Donetsk National University official group in the period from January to March 2019 has been analyzed. The average coefficient of attraction of users for three types of publications is defined. The main technologies of astroturfing that can be implemented in social networks for creation of a negative image of a higher education institution are represented. There are: custom comments and «posts» on Internet resources are considered; ordered trolling; creation of pages – «clones» or pages –«fake»; placing «viral» photographs, audio and video materials in public access and distribution among users of the social network.

## References

1. Azoury, N., Daou, L., Khoury, C. E.: University image and its relationship to student satisfaction- case of the Middle Eastern private business schools. International Strategic Management Review 2, 1–8 (2014).
2. Batychko, G. I., Veliyeva, O. R.: Social networks as a forming factor of positive image and positioning space of the university at the educational service market. Bulletin of the Mariupol State University. Series: Philosophy, Culturology, Sociology 3, 17–25 (2012).
3. Danko, Yu. A.: Asturfing as an instrument of virtual manipulation and political propaganda in the information age. Modern society: political sciences, sociological sciences, cultural sciences 2, 38–49 (2015).
4. Facebook. Faculty of Mathematics and IT of Vasyl Stus DonNU. URL: https://www.facebook.com/math.donnu/ last accessed 2019/04/15.
5. Fadeeva, M. V.: Psychological conditions of training of heads of general educational institutions. author's abstract dis. for the sciences. Degree Candidate ps. sciences. Kyiv (2010).
6. Gorbatiuk, T. V.: The development of Internet: perspective and prospects. Scientific Bulletin of NUBiP of Ukraine. Series: Humanities Studies 274, 44–51 (2017).
7. Holland, D., Krause, A., Provencher, J., Seltzer, T.: Transparency tested: The influence of message features on public perceptions of organizational transparency. Public Relations Review. 44(2), 256–264 (2018).
8. Horbenko, N. V.: Features of shaping the image of a modern university. Educational discourse 1(5), 36–43 (2014).
9. Kirichok, A.: The use of new media in shaping the image of universities. Bulletin of the Book Chambe 2, 42–44 (2015).
10. Korzh, R., Fedushko, S., Peleschyshyn, A.: Methods for forming an informational image of a higher education institution. Webology 12(2), 1–10 (2015). http://www.webology.org/2015/v12n2/a140.pdf
11. Korzh, R., Peleshchyshyn, A., Fedushko, S., Syrov, Y.: Protection of University Information Image from Focused Aggressive Actions. Advances in Intelligent Systems and Computing: Recent Advances in Systems, Control and Information Technology, SCIT 2016, 543, pp. 104–110. Springer, Poland (2017). DOI: 10.1007/978-3-319-48923-0_14.
12. Korzh, R., Peleshchyshyn, A., Syrov, Yu., Fedushko, S.: University's Information Image as a Result of University Web Communities' Activities. Advances in Intelligent Systems and Computing: Selected Papers from the International Conference on Computer Science

and Information Technologies, CSIT 2016, 512, pp. 115-127. Springer, Ukraine (2017). DOI: 10.1007/978-3-319-45991-2_8.
13. Number of internet users worldwide from 2005 to 2018 (in millions). https://www.statista.com/statistics/273018/number-of-internet-users-worldwide/. last accessed 2019/04/12.
14. Number of social media users worldwide from 2010 to 2021 (in billions). https://www.statista.com/statistics/278414/number-of-worldwide-social-network-users.
15. Pavlov, S. N.: Information factor in formation of image of higher educational institution. Fundamental research 9, 635–640 (2012).
16. Pavlov, S. N.: Main principles of the conception of the effective image formation of the higher educational institution. Fundamental research 4, 1216–1221 (2013).
17. Peng, J., Detchon, S., Choo, K. R.: Astroturfing detection in social media: a binary ngram–based approach. Concurrency and Computation. Practice and Experience 29(17), (2016).
18. Petrenko, O. S.: Specificity of Technologies for Influence on Public Opinion in Electronic Social Networks and Social Media. Modern social problems in the measurement of sociology of management: a collection of scientific works of the DonNUU 281, 241–249 (2014).
19. Pyramid of world population from 1950 to 2100, https://www.populationpyramid.net/ru/%D0%BC%D0%B8%D1%80-%D0%B7%D0%B5%D0%BC%D0%BB%D1%8F/2015/, last accessed 2019/04/12.
20. Rassokhin, V. V., Chaprak, N. V.: Assessment of the investment market using the relative range of asset price fluctuations. Finance and credit 29, 13–28 (2015).
21. Sadigora, T. S.: Social and psychological functions of social networks. TSU Science Vector 3, 192–194 (2012).
22. Social networks in 2018: a global study, https://www.web-canape.ru/business/socialnye-seti-v-2018-godu-globalnoe-issledovanie/, last accessed 2019/04/19.
23. Toda, A., Carmo, R., Silva, A., Bittencourt, I., Isotani, S.: An approach for planning and deploying gamification concepts with social networks within educational contexts. International Journal of Information Management 46, 294–303 (2019).
24. Tsykhovska, E. D.: Self-presentation in social networks: Facebook account as a tool of image creation. Dnipropetrovsk University Bulletin 17, 137–147 (2017).
25. Vitoropoulou, M., Karyotis, V., Papavassiliou, S.: Sensing and monitoring of information diffusion in complex online social networks. Peer-to-Peer Networking and Applications 12(3), 604–619 (2018).
26. What is the engagement rate and how to calculate it? https://geniusmarketing.me/lab/koefficient-vovlechennosti/ last accessed 2019/04/16.
27. What social networks are popular among residents of Ukraine: statistics by year, https://tech.informator.ua/2018/11/01/kakie-sotsseti-populyarny-u-zhitelej-ukrainy-statistika-po-godam/, last accessed 2019/04/14.